\documentclass[%
amsmath,amssymb,
twocolumn,
]{revtex4-1}

\usepackage{graphicx}
\usepackage{dcolumn}
\usepackage{bm}
\usepackage{xcolor}
\usepackage[normalem]{ulem}
\usepackage{amsmath}
\usepackage{algorithm}
\usepackage[noend]{algpseudocode}
\usepackage{enumitem}

\begin{document}

\title{Phonon-Phonon Interactions in Strongly Bonded Solids:  Selection Rules and Higher-Order Processes}

\author{Navaneetha K. Ravichandran}
\email{navaneeth@iisc.ac.in}
\affiliation{Department of Mechanical Engineering, Indian Institute of Science, Bangalore 560012, India}
\author{David Broido}
\affiliation{Department of Physics, Boston College, Chestnut Hill, MA 02467, USA}


\begin{abstract}
We show that the commonly used lowest-order theory of phonon-phonon interactions frequently fails to accurately describe the anharmonic phonon decay rates and thermal conductivity ($\kappa$), even among strongly bonded crystals. Applying a first principles theory that includes both the lowest-order three-phonon and the higher-order four-phonon processes to seventeen zinc blende semiconductors, we find that the lowest-order theory drastically overestimates the measured $\kappa$ for many of these materials, while inclusion of four-phonon scattering gives significantly improved agreement with measurements. We have identified new selection rules on three-phonon processes that help explain many of these failures in terms of anomalously weak anharmonic phonon decay rates predicted by the lowest-order theory competing with four-phonon processes. We also show that zinc blende compounds containing boron (B), carbon (C) or nitrogen (N) atoms have exceptionally weak four-phonon scattering, much weaker than in compounds that do not contain B, C or N atoms. This new understanding helps explain the ultrahigh $\kappa$ in several technologically important materials like cubic boron arsenide, boron phosphide and silicon carbide. At the same time, it not only makes the possibility of achieving high $\kappa$ in materials without B, C or N atoms unlikely, but it also suggests that it may be necessary to include four-phonon processes in many future studies. Our work gives new insights into the nature of anharmonic processes in solids and demonstrates the broad importance of higher-order phonon-phonon interactions in assessing the thermal properties of materials. 
\end{abstract}
\maketitle
\section{Introduction}
Phonon-phonon interactions arise from the inherent anharmonicity of the interatomic bonds in solids.  They determine fundamental properties of crystals such as the lattice thermal conductivity ($\kappa$) as well as the infra-red, Raman and neutron scattering cross sections~\cite{peierls_zur_1929_ref1, maradudin_scattering_1962, cowley_anharmonic_1968}. They impact diverse phenomena including phonon drag~\cite{geballe_seebeck_1954}, phonon bottleneck~\cite{fu_hot_2017, yang_acoustic-optical_2017} and hydrodynamic thermal transport~\cite{pohl_observation_1976, huberman_observation_2019}.\\

In his pioneering work, Peierls described intrinsic thermal resistance in semiconductors and insulators using the lowest-order theory of phonon-phonon interactions, which involves three phonons~\cite{peierls_zur_1929_ref1}. The complexity of this theory forced simple model approximations to be used until the past decade, when these simple models have been supplanted by first principles computational approaches to describe anharmonic phonon decay and heat conduction~\cite{lindsay_survey_2018_ref2, mcgaughey_phonon_2019_ref3, lindsay_perspective_2019_ref4}, still based on the lowest-order theory.\\

While this lowest-order theory has been generally assumed to adequately describe anharmonic properties of crystals, \textsl{ab initio} descriptions of higher-order four-phonon interactions are now becoming possible~\cite{feng_quantum_2016_ref5, feng_four-phonon_2017_ref6, ravichandran_unified_2018_ref7, xia_revisiting_2018_ref8, xia_anharmonic_2018_ref9, tian_unusual_2018_ref19, chen_ultrahigh_2020, ravichandran_non-monotonic_2019, yang_stronger_2019, mellan_electron_2019}. One might expect higher-order phonon-phonon interactions to be important in strongly anharmonic materials, i.e., those with weak chemical bonds.  Indeed, recent work has shown that for such materials, not only are four-phonon processes important~\cite{ravichandran_unified_2018_ref7, xia_revisiting_2018_ref8, xia_anharmonic_2018_ref9}, but the anharmonic renormalization of phonon modes themselves~\cite{ravichandran_unified_2018_ref7, xia_revisiting_2018_ref8, xia_anharmonic_2018_ref9, hellman_lattice_2011_ref10, hellman_temperature_2013_ref11, errea_first-principles_2013_ref12, souvatzis_entropy_2008_ref13, van_roekeghem_anomalous_2016_ref14, ribeiro_strong_2018_ref15, tadano_self-consistent_2015_ref16, tadano_first-principles_2018_ref17, allen_anharmonic_2015_ref18} is required to accurately match measured phonon frequency, thermal expansion and thermal conductivity data.\\
 
But what about weakly anharmonic materials i.e. those with relatively strong interatomic bonding?  How well does the lowest-order theory work for such cases?  Significant failures of the lowest-order theory have been found recently in cubic boron arsenide (BAs)~\cite{feng_four-phonon_2017_ref6, tian_unusual_2018_ref19}, boron antimonide (BSb)~\cite{ravichandran_non-monotonic_2019} and aluminum antimonide (AlSb)~\cite{yang_stronger_2019}.  But, the underlying reasons for the failures in BAs and BSb apply to very few materials, and a full explanation for the failure in AlSb has not been given in Ref.~\cite{yang_stronger_2019}, as described below.  At the same time, the lowest-order theory has been found to work well for diamond~\cite{feng_four-phonon_2017_ref6, ravichandran_unified_2018_ref7}, gallium nitride (GaN)~\cite{yang_stronger_2019} and cubic boron nitride (BN)~\cite{chen_ultrahigh_2020}. The computational cost of four-phonon calculations has thus far restricted consideration to only this small number of weakly anharmonic compounds.  As a result, To date, a number of fundamental questions about the nature of anharmonic phonon-phonon processes in solids remain unanswered.  Specifically:  Are the failures of the lowest-order theory of phonon-phonon interactions rare or common among weakly anharmonic materials? Is there a general framework to understand the interplay between three-phonon and four-phonon processes that can drive the failures?  How do four-phonon scattering strengths vary across such materials?  Can new trends be identified to aid in the search for materials with high $\kappa$? \\

\begin{figure}[!ht]
\begin{center}
\includegraphics*[scale=0.4]{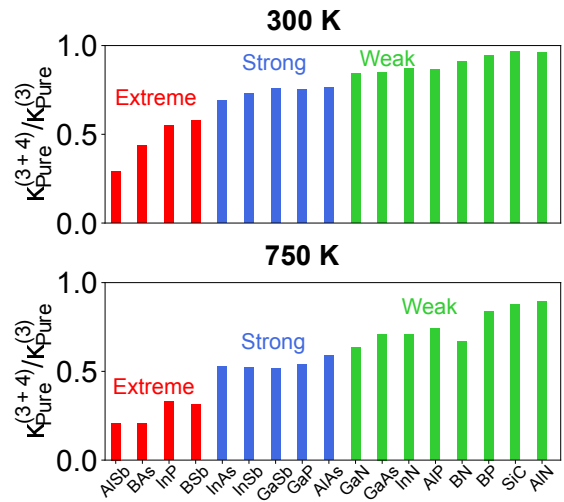}
\end{center}
\caption{Ratios of the thermal conductivities including four-phonon scattering to those omitting it ($\kappa^{(3+4)}_{\mathrm{Pure}}/\kappa^{(3)}_{\mathrm{Pure}})$ for seventeen zinc blende compounds at 300 K (top panel) and 750 K (bottom panel).  The large thermal conductivity reductions even at 300 K show that the lowest-order theory of phonon-phonon interactions is failing for many of these materials.}	\label{KappaHistogram_16Mat_InkscapeEdited_v2_Q51}
\end{figure}

To address these questions, in this work, we examine the anharmonic phonon decay rates and thermal conductivities of seventeen different zinc blende (ZB) semiconductors with widely varying phonon properties using a first principles computational approach that includes both the lowest-order three-phonon and higher-order four-phonon interactions as well as interactions between phonons and the isotopic mass disorder found in most crystals. The findings for $\kappa$ are summarized in Fig.~\ref{KappaHistogram_16Mat_InkscapeEdited_v2_Q51} and confirmed by the available measured data for these materials (shown later). It is evident from the figure that failures of the lowest-order theory are remarkably common among ZB compounds, with nine of them showing a reduction in $\kappa$ of at least 20\% at room temperature, and at least 40\% at 750 K. For many of the seventeen compounds, anharmonic decay rates of the optic phonons from four-phonon scattering are as large or even larger than those predicted by the lowest-order three-phonon theory. Along with the previously identified compounds, BAs, BSb and AlSb, we find a particularly catastrophic failure of the lowest-order theory to occur for indium phosphide (InP). Yet, the properties responsible for the failures in BAs and BSb cannot explain the behavior in InP, as we show below.  The same is true of the significant failures seen in gallium antimonide (GaSb), indium arsenide (InAs) and indium antimonide (InSb). At the same time, the lowest-order theory works quite well for many compounds containing boron, carbon or nitrogen atoms.\\ 

To fundamentally understand these findings, we first identify a set of selection rules that follow from energy and quasi-momentum conservation on the lowest-order three-phonon scattering and are dictated entirely by features in the phonon dispersions. To the extent that a material's phonon dispersions activate these selection rules, the phase space for the corresponding lowest-order three-phonon scattering channels are weakened or frozen out, which can result in unusually long lifetimes for both acoustic and optic phonons with correspondingly large contributions to $\kappa$, predicted in the lowest-order theory.  While some of these selection rules have been identified previously~\cite{lax_spontaneous_1981_ref25, lindsay_first-principles_2013_ref20, mukhopadhyay_optic_2016_ref21}, they cannot explain the behavior found in many of the studied compounds.  We have identified new selection rules that are critical for developing a full understanding of the failures of the lowest-order theory in the zinc blende compounds.\\
 
Next, we show that in striking contrast, selection rules do not affect the four-phonon scattering rates, which generally increase with frequency for all seventeen materials. Thus, whenever the selection rules on the lowest-order processes drive weak three-phonon scattering, four-phonon scattering can become more important. Most remarkably, the materials naturally divide themselves into two groups according to the strength of the four-phonon scattering processes. In the materials not containing boron, carbon or nitrogen (non-BCN compounds) the four-phonon scattering rates are of similar strength. For materials containing boron, carbon or nitrogen (BCN compounds) four-phonon scattering rates are up to an order of magnitude smaller than those in the non-BCN compounds, a difference we trace to the unusually strong bonding in the BCN compounds. As expanded on below, the large difference in the strength of four-phonon scattering in the non-BCN compounds compared with the BCN compounds has profound implications on the phonon lifetimes and $\kappa$.

\section{Selection Rules on Phonon-Phonon Processes}
We begin by describing six selection rules for the lowest-order phonon-phonon scattering and discuss their interplay with the higher-order interactions.  Some of these selection rules have been discussed previously~\cite{lax_spontaneous_1981_ref25, lindsay_first-principles_2013_ref20, mukhopadhyay_optic_2016_ref21}.  We focus on crystals with two different atoms in the unit cell, which encompasses a large number of semiconductors and insulators including the technologically important zinc blende compounds studied in the present work, which have inherently weak anharmonicity of the crystal bonds. This fact is demonstrated by the small thermal expansion and weak effects of anharmonic renormalization on their phonon frequencies (see Supplementary Materials [SM] section S1 for phonon dispersions and SM section S2 for details on thermal expansion for all seventeen compounds).   At the same time, the breadth of the considered materials provides simple yet sufficiently different phonon dispersions that enable exploration of the effects of various selection rules, co-existing to different extents in different materials, on their anharmonic phonon decay rates and thermal conductivities.\\

\subsection{Lowest-Order Processes}
We divide the six phonon polarizations (branches) into three acoustic (A) and three optic (O) branches.  There are four possible combinations of the lowest-order anharmonic processes that involve the decay of a phonon into two others or the coalescence of two phonons into a third. These are:  (i)  (AAA); (ii)   (AAO);  (iii)   (AOO); (iv)   (OOO).  These processes must satisfy conservation of phonon energy and quasi-momentum~\cite{ziman1960electrons_ref26}:
\begin{align}
\omega_j\left(\mathbf{q}\right) \pm \omega_{j'}\left(\mathbf{q}'\right) &= \omega_{j''}\left(\mathbf{q}'' + \mathbf{G}\right)		\label{econs}\\
\mathbf{q} \pm \mathbf{q}'' &= \mathbf{q}'' + \mathbf{G}	\label{qcons}
\end{align}
Here, the three phonons participating in the scattering process have wave vectors $\mathbf{q}$, $\mathbf{q}'$ and $\mathbf{q}''$, branches, $j$, $j'$ and $j''$, and frequencies, $\omega_j\left(\mathbf{q}\right)$, $\omega_{j'}\left(\mathbf{q}'\right)$ and $\omega_{j''}\left(\mathbf{q}''\right)$ respectively, and $\mathbf{G}$ is a reciprocal lattice vector to account for normal ($\mathbf{G} = 0$) and Umklapp ($\mathbf{G} \ne 0$) processes.\\
 
For each phonon mode, $\left(j, \mathbf{q}\right)$, and fixed $j'$, $j''$, the phase space for the lowest-order scattering is defined by the set of wavevectors - $\mathbf{q}'$ and $\mathbf{q}''$, that satisfy Eqs.~\ref{econs} and~\ref{qcons}~\cite{lindsay_survey_2018_ref2, mcgaughey_phonon_2019_ref3, lindsay_three-phonon_2008_ref22, li_shengbte:_2014_ref23, lee_resonant_2014_ref24}. Features in the phonon dispersions that cause the phase space to vanish or significantly weaken define selection rules on these processes.\\
 
First, we note a selection rule on OOO processes:  energy conservation freezes out OOO processes if the optic phonon bandwidth, $\Delta\omega_O$, is smaller than the lowest optic phonon frequency.  This is the case in all compounds studied in the present work, so we will not discuss the OOO selection rule further.  We have identified five selection rules on AAA, AAO, and AOO processes that follow directly from Eqs.~\ref{econs} and~\ref{qcons}, combining those previously found~\cite{lax_spontaneous_1981_ref25, lindsay_first-principles_2013_ref20, mukhopadhyay_optic_2016_ref21} with two new ones. These selection rules divide into two types that complement each other.  The first two selection rules (classified as Type 1 selection rules) follow directly from energy conservation (Eq.~\ref{econs}):  
\subsubsection{Type 1 selection rules}
\begin{enumerate}
\item \textbf{AOO\#1 selection rule}: AOO processes cannot occur involving acoustic phonons whose frequencies exceed the optic phonon bandwidth. This selection rule establishes a cut-off frequency ($\Delta\omega_O$) in the acoustic phonon spectrum beyond which an acoustic phonon cannot participate in AOO processes. This cut-off has been pointed out previously~\cite{mukhopadhyay_optic_2016_ref21}.
\item \textbf{AAO selection rule}: An AAO process cannot occur if one of the participating acoustic phonons has a frequency smaller than the frequency gap between acoustic and optic phonons. This selection rule establishes a cut-off frequency for the acoustic phonons, $\Delta\omega_{A-O}$ - the A-O frequency gap, below which an acoustic phonon cannot participate in AAO processes. It also establishes a cut-off frequency for the optic phonons, $2\Delta\omega_A$ ($\Delta\omega_A$ is the acoustic phonon bandwidth), above which optic phonons cannot participate in AAO processes. From this selection rule, it follows that for materials with A-O gaps larger than the highest acoustic phonon frequency, $\Delta\omega_A$, AAO processes are completely forbidden~\cite{lindsay_first-principles_2013_ref20}.
\end{enumerate}
If $\Delta\omega_O < \Delta\omega_{A-O}$, the AAO and AOO\#1 selection rules act in unison to create a frequency window for the acoustic phonons where both AAO and AOO scattering channels are completely forbidden, and only the AAA three-phonon scattering channel is allowed. This is illustrated schematically in Figs.~\ref{Fig2_Schematic_v3} (b) and (d).  Similarly, if $\Delta\omega_O > \Delta\omega_A - \Delta\omega_{A-O}$, then a frequency window is created where only AOO scattering by optic phonons can occur.  This is illustrated in Fig.~\ref{Fig2_Schematic_v3} (e). If frequency windows are created by the AAO and AOO\#1 selection rules, then the following three selection rules (classified as Type 2 selection rules) can drive anomalously weak three-phonon scattering:
\subsubsection{Type 2 selection rules}
\begin{enumerate}[resume]
\item \textbf{AAA\#1 selection rule}: \textsl{An anharmonic process of any order in which one phonon decays into a set of other phonons with higher phase velocity cannot occur}. This selection rule was proven by Lax, Hu and Narayanamurti~\cite{lax_spontaneous_1981_ref25}.
\item \textbf{AAA\#2 selection rule}: \textsl{As the group velocity of longitudinal acoustic (LA) phonons increases compared with transverse acoustic (TA) phonon velocities, the phase space of low frequency AAA processes decreases}. To our knowledge, this selection rule has not been pointed out before. Unlike the other selection rules, this one does not completely forbid any particular three-phonon scattering channel except in the limit of infinite LA phonon velocities. However, the trend it expresses is critical in explaining the anomalously small acoustic three-phonon scattering rates in several zinc blende compounds discussed in the next section. This selection rule is examined in more detail in the SM section S8(B).
\item \textbf{AOO\#2 selection rule}: \textsl{An AOO process is forbidden when both optic phonons derive from the same phonon branch, provided the group velocities of the optic phonons are smaller than those of the acoustic phonons}. To our knowledge, this selection rule has not been pointed out before. Typically, the group velocities of optic phonons are smaller than those of the acoustic phonons  that can participate in an AOO process while satisfying energy conservation. We have confirmed numerically that this selection rule is satisfied for all seventeen materials in this study. Exceptions can occur in materials with larger optic phonon group velocities, such as the lead chalcogenides. Further discussion is given in the SM section S8(A).
\end{enumerate}

\subsection{Minimal Influence of Selection Rules on Four-Phonon Processes}
Ziman has suggested that selection rules on four-phonon processes are less restrictive than those on three-phonon processes~\cite{ziman1960electrons_ref26}. Selection rules do exist for some four-phonon processes.  For example, the general selection rule noted in Ref.~\cite{lax_spontaneous_1981_ref25} forbids $A\leftrightarrow A+A+A$ processes when all phonons derive from the same branch. However, it does not forbid the corresponding $A + A\leftrightarrow A+A$ processes.  Similarly, processes such as $A\leftrightarrow A+O+O$ do not conserve energy, while a large phase space exists for $A + O\leftrightarrow A+O$ (AOAO) processes.  Our quantitative first principles calculations show that the AOAO processes dominate in all seventeen compounds studied in this work, with AAAA and OOOO processes also being important for acoustic and optic phonons respectively, and the influence of selection rules on four-phonon processes is minimal.\\

\subsection{Interplay of lowest-order selection rules and four-phonon scattering}
\begin{figure}[h]
\begin{center}
\includegraphics*[scale=0.335]{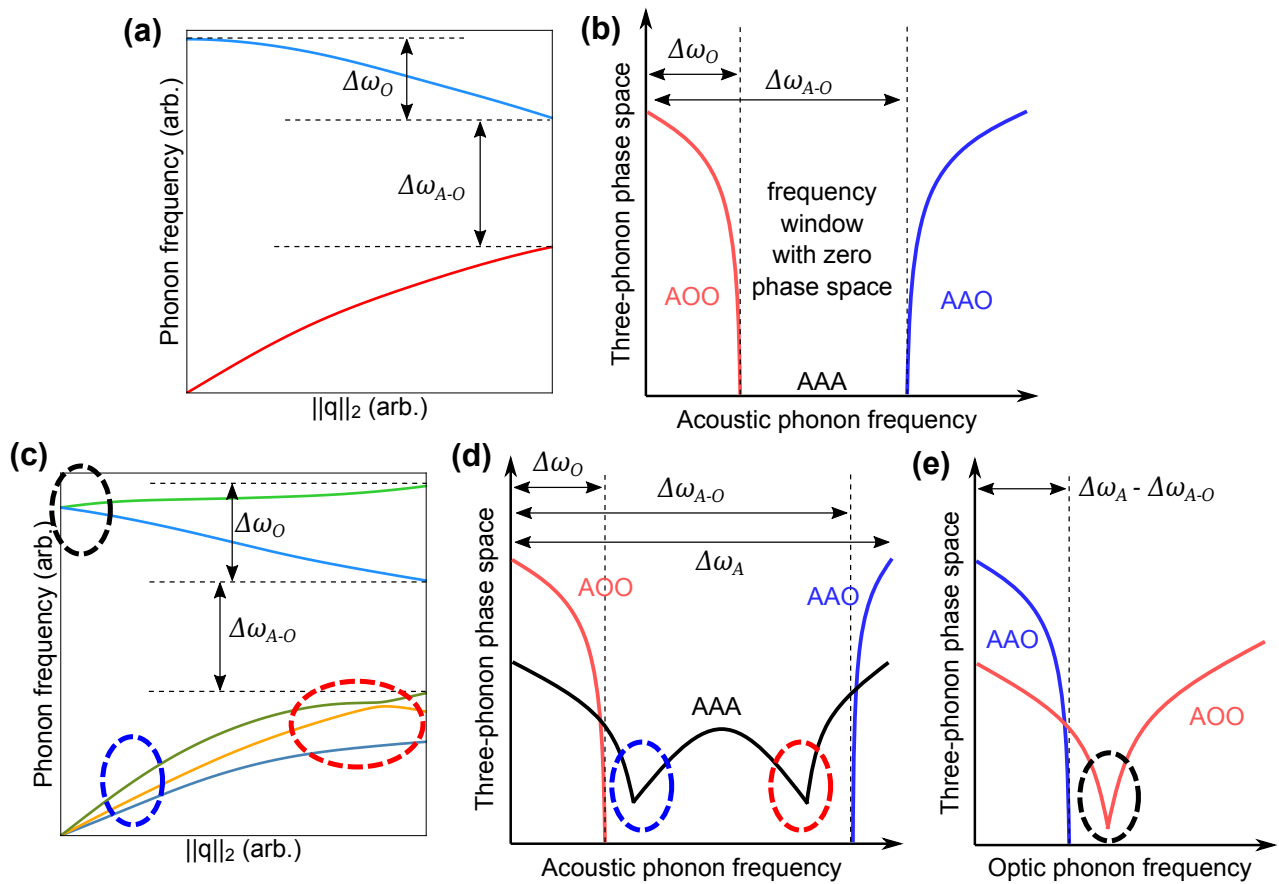}
\end{center}
\caption{Hypothetical phonon dispersions (a) and the corresponding three-phonon phase space (b), where degenerate acoustic phonons, small optic phonon bandwidth ($\Delta\omega_O$) and large A-O frequency gap ($\Delta\omega_{A-O}$) result in a frequency window for the acoustic phonons without any phase space for three-phonon scattering (due to AAO, AOO\#1 and AAA\#1 selection rules). Also shown are the features of the dispersions of a real material (c), where acoustic phonons having large LA-TA velocity ratio (blue dashed oval) and/or being \textsl{bunched together} (red dashed oval), along with small $\Delta\omega_O$ and large A-O gap result in a frequency window with weak total three-phonon scattering phase space due to AAO, AOO\#1, AAA\#1 and AAA\#2 selection rules (d). Furthermore, near-degeneracy in the optic phonons (black dashed oval in (c), (e)), along with a large A-O gap can result in weak three-phonon scattering phase space for the optic phonons due AAO and AOO\#2 selection rules. The curves labeled AAA, AAO and AOO in Figs. (b), (d) and (e) show schematically the regions in which the corresponding three-phonon processes occur.}	\label{Fig2_Schematic_v3}
\end{figure}

The above selection rules impose strict constraints on the phase space for specific three-phonon processes, which are governed completely by a material's phonon dispersions – a harmonic quantity. As described earlier, the selection rules complement each other: For materials whose constituent atoms have different masses (so as to create finite $\Delta\omega_{A-O}$) and relatively low ionicity, (so that $\Delta\omega_O$ is relatively small), AAO and AOO\#1 selection rules create frequency windows where only AAA scattering can occur for acoustic phonons, and only AOO scattering can occur for optic phonons. Then, if features in the material's phonon dispersions activate AAA\#1, AAA\#2 or AOO\#2 selection rules, regions of small total three-phonon phase space result. In principle, if hypothetical phonon dispersions existed with a collection of features described in the above selection rules, then \textsl{the lowest-order theory would predict infinite intrinsic lifetimes and dissipationless transport for the affected phonon modes}. For example, if the three acoustic phonon branches of hypothetical phonon dispersions coincided throughout the Brillouin zone (BZ), then the AAA\#1 selection rule requires that the phase space for AAA scattering vanish.  Then, if $\Delta\omega_{A-O} > \Delta\omega_O$, a frequency window would be created with zero phase space for lowest-order phonon-phonon scattering for acoustic phonons, according to AOO\#1 and AAA\#1 selection rules. This scenario is illustrated in Figs.~\ref{Fig2_Schematic_v3} (a) and (b). Similarly, if all of the optic phonon branches of hypothetical phonon dispersions coincided throughout the BZ, the optic phonon velocities were much smaller than those of the acoustic phonons available that satisfy energy conservation, and $\Delta\omega_{A-O} > \Delta\omega_A$, then the optic phonons could not decay at all due to AAO and AOO\#2 selection rules (no separate plot is shown for the optic phonon phase space, since it would be uniformly zero in this scenario). Although these scenarios do not occur in real materials, insofar as features in the real dispersions approach those in the hypothetical ones, selection rules will be activated that reduce the phase space for the lowest-order processes. For example, if acoustic branches have similar frequencies in some region of the BZ, then the phase space for some AAA processes involving phonons with frequencies in this region becomes small. If AAO and AOO\#1 selection rules create a frequency window where the AAA processes are weak, the total three-phonon scattering phase space will be small, as shown in Fig.~\ref{Fig2_Schematic_v3} (d). This is what happens in BAs and BSb, as has been explained previously~\cite{lindsay_first-principles_2013_ref20}. Similarly, if phonons from different optic branches become nearly degenerate in some region of the BZ and their frequencies are larger than $\Delta\omega_A - \Delta\omega_{A-O}$, then the phase space for the corresponding AOO processes becomes weak. This is illustrated in Fig.~\ref{Fig2_Schematic_v3} (e). We will show in the next section, that this scenario explains the large contributions to the thermal conductivity from optic phonons in BAs and AlSb. We note that the degree of \textsl{activation} of the \textsl{type 2} selection rules is qualitative and cannot be fully assessed by visual inspection of phonon dispersions. Quantitative \textsl{ab initio} calculations are required to determine the impact of the selection rules on the anharmonic decay rates and thermal conductivity of a given material.
 
Higher-order four-phonon scattering rates are generally weaker than their lowest-order counterparts in weakly anharmonic materials at low-to-moderate temperatures.  However, here we emphasize that when features in the phonon dispersions cause the above-described selection rules to impose severe constraints on the phase space for the lowest-order phonon-phonon processes, the effects of four-phonon scattering can become quite important. In the next section and in the SM section S3, the calculated thermal conductivities of the studied seventeen different materials are presented and compared to measured data, and the failures and successes of the lowest-order theory are explained in terms of the selection rules and their interplay with the varying strengths of four-phonon scattering.\\

\onecolumngrid

\twocolumngrid

\section{Results}
\begin{figure}[h]
\begin{center}
\includegraphics*[scale=0.29]{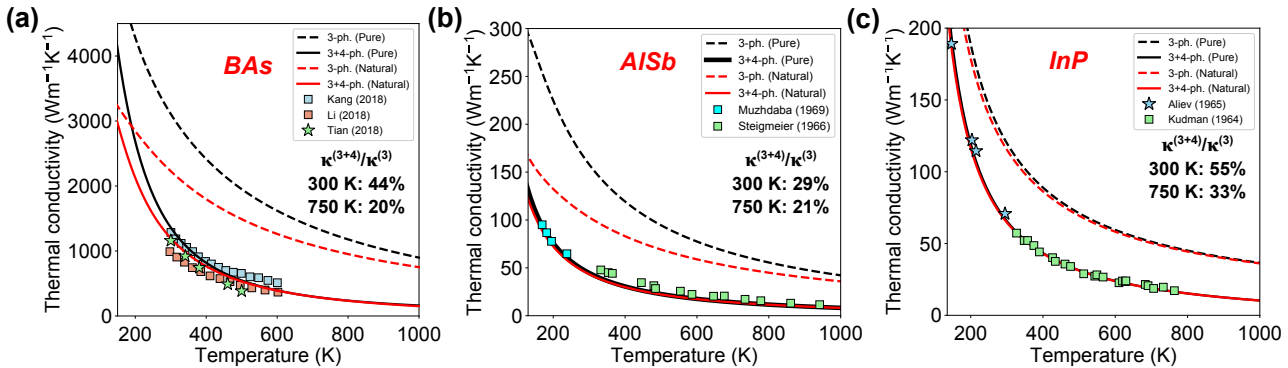}
\end{center}
\caption{First principles computed $\kappa^{(3)}_{\mathrm{Pure}}$ (dashed black), $\kappa^{(3+4)}_{\mathrm{Pure}}$ (solid black), $\kappa^{(3)}_{\mathrm{Nat.}}$ (dashed red) and $\kappa^{(3+4)}_{\mathrm{Nat.}}$ (solid red) for BAs (a), AlSb (b) and InP (c) as functions of temperature and compared with the experimental data in the literature. The experimental measurements of $\kappa$ are from (a) BAs: \textsl{Kang et al.}~\cite{kang_experimental_2018_ref27}, \textsl{Li et al.}~\cite{li_high_2018_ref28} and \textsl{Tian et al.}~\cite{tian_unusual_2018_ref19} (b) AlSb: \textsl{Muzhdaba et al.}~\cite{muzhdaba1969thermal_ref29} and \textsl{Steigmeier et al.}~\cite{steigmeier_acoustical-optical_1966_ref30_34} (c) InP: \textsl{Aliev et al.}~\cite{aliev1965thermal_ref31} and \textsl{Kudman et al.}~\cite{kudman_thermal_1964_ref32}.}	\label{Fig2_HighKappaMat_kvsT_WithExpt}
\end{figure}

\begin{figure}[h]
\begin{center}
\includegraphics*[scale=0.28]{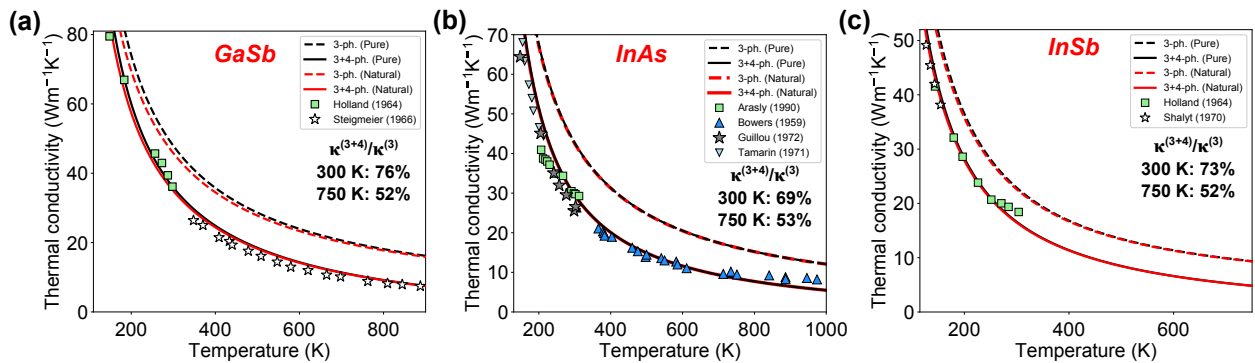}
\end{center}
\caption{First principles computed $\kappa^{(3)}_{\mathrm{Pure}}$ (dashed black), $\kappa^{(3+4)}_{\mathrm{Pure}}$ (solid black), $\kappa^{(3)}_{\mathrm{Nat.}}$ (dashed red) and $\kappa^{(3+4)}_{\mathrm{Nat.}}$ (solid red) for GaSb (a), InAs (b) and InSb (c) as functions of temperature and compared with the experimental data in the literature. The experimental measurements of $\kappa$ are from (a) GaSb: \textsl{Holland et al.}~\cite{holland_phonon_1964_ref33_39} and \textsl{Steigmeier et al.}~\cite{steigmeier_acoustical-optical_1966_ref30_34} (b) InAs: \textsl{Arasly et al.}~\cite{arasly1990characteristics_ref35}, \textsl{Bowers et al.}~\cite{bowers_inas1xpx_1959_ref36}, \textsl{Guillou et al.}~\cite{le_guillou_phonon_1972_ref37} and \textsl{Tamarin et al.}~\cite{tamarin1971thermal_ref38} (c) InSb: \textsl{Holland et al.}~\cite{holland_phonon_1964_ref33_39} and \textsl{Shalyt et al.}~\cite{shalyt_scattering_1970_ref40}.}	\label{Fig4_MediumKappaMat_kvsT_WithExpt}
\end{figure}

The thermal conductivities of the seventeen compounds with zinc blende structure - GaAs, AlP, AlAs, GaP, GaSb, InAs, InSb, AlSb, InP, BN, BP, BAs, BSb, SiC, GaN, InN and AlN - are obtained as functions of temperature by solving the Boltzmann equation for phonon transport including three-phonon, four-phonon and phonon-isotope scattering using a recently-developed first principles approach~\cite{ravichandran_unified_2018_ref7}. A brief description of this approach is included in the Methods section and Appendix.  Of  the seventeen materials considered here, all but five (AlP, GaN, InN, AlN and BSb) have measured thermal conductivity data available for comparison. Six such comparisons are presented in Fig.~\ref{Fig2_HighKappaMat_kvsT_WithExpt} for BAs, AlSb and InP and Fig.~\ref{Fig4_MediumKappaMat_kvsT_WithExpt} for GaSb, InAs and InSb, while the remaining (except BN) are given in the SM section S3. The case of BN has been extensively described in the main text and the supplementary information of our recent publication~\cite{chen_ultrahigh_2020}, so we do not present those plots in the SM. Calculated thermal conductivities are given for the following four conditions: lowest-order three-phonon scattering only ($\kappa^{(3)}_{\mathrm{Pure}}$), three-phonon scattering and phonon scattering from the natural isotopic disorder ($\kappa^{(3)}_{\mathrm{Nat.}}$), three-phonon and four-phonon scattering only ($\kappa^{(3+4)}_{\mathrm{Pure}}$), and three-phonon, four-phonon and phonon-isotope scattering ($\kappa^{(3+4)}_{\mathrm{Nat.}}$). Calculated RT $\kappa$ values and convergence studies for each material are tabulated in SM Tables II and III.  In many of the cases, $\kappa\left(T\right)$ calculated from the lowest-order theory significantly overestimates the measured $\kappa\left(T\right)$, while including four-phonon interactions brings the calculated $\kappa\left(T\right)$ curves into very good agreement with the measured data~\cite{chen_ultrahigh_2020, tian_unusual_2018_ref19, kang_experimental_2018_ref27, li_high_2018_ref28, muzhdaba1969thermal_ref29, steigmeier_acoustical-optical_1966_ref30_34, aliev1965thermal_ref31, kudman_thermal_1964_ref32, holland_phonon_1964_ref33_39, arasly1990characteristics_ref35, bowers_inas1xpx_1959_ref36, le_guillou_phonon_1972_ref37, tamarin1971thermal_ref38, shalyt_scattering_1970_ref40}. As already seen in Fig.~\ref{KappaHistogram_16Mat_InkscapeEdited_v2_Q51}, four-phonon scattering suppresses the $\kappa^{(3)}_{\mathrm{Pure}}$ for nine of the seventeen compounds by at least 20\% at room temperature (RT, 300K) and at least 40\% at 750K.  The larger suppressions at 750K are explained by the more rapid increase of four-phonon scattering rates with temperature compared with three-phonon scattering rates~\cite{feng_four-phonon_2017_ref6, chen_ultrahigh_2020}. Thus, it is clear that the lowest-order theory is failing for many of these compounds. \\

\onecolumngrid

\twocolumngrid

As explained in the previous section, activation of the Type 1 selection rules can create frequency windows where only AAA can occur for acoustic phonons and only AOO processes can occur for optic phonons. Pockets of small three-phonon scattering phase space result if Type 2 selection rules are simultaneously activated.  The most significant effects from this combined activation of Type 1 and Type 2 selection rules occurs in the four compounds - BAs, BSb, AlSb and InP, those falling in the “extreme” classification in Fig.~\ref{KappaHistogram_16Mat_InkscapeEdited_v2_Q51}. Since different Type 2 selection rules are required to explain the extreme behavior in these materials, in the results discussed below materials are grouped according to which of the Type 2 selection rules (AAA\#1, AAA\#2 and AOO\#2) are activated. We note that the degree of activation of Type 2 selection rules is somewhat qualitative but will be evident from the three-phonon phase space plots.\\

\subsection{Materials influenced by AAA\#1 selection rule: BAs, BSb, GaN, BP, BN, SiC, AlN}
\begin{figure}[h]
\begin{center}
\includegraphics*[scale=0.4]{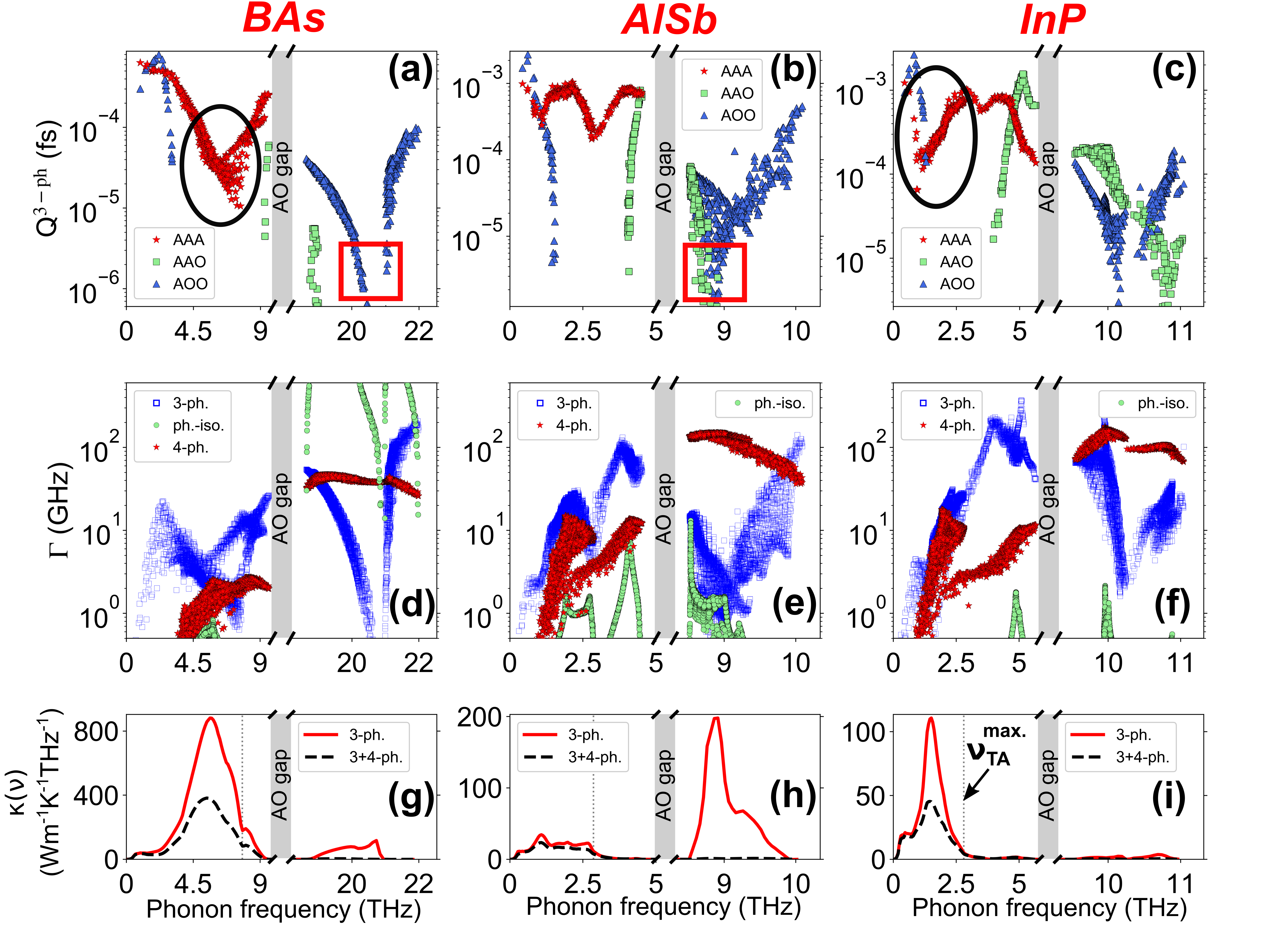}
\end{center}
\caption{Calculated three-phonon scattering phase space for different scattering channels (a)-(c), three-phonon, four-phonon and phonon-isotope scattering rates at 300 K (d)-(f) and spectral contributions to $\kappa^{(3)}_{\mathrm{Pure}}$ (solid red) and $\kappa^{(3+4)}_{\mathrm{Pure}}$ (dashed black) at 300 K (g)-(i), for BAs (first column: (a), (d) and (g)), AlSb (second column: (b), (e) and (h)) and InP (third column (c), (f) and (i)). The dotted vertical lines in (g)-(i) show the location of the maximum TA phonon frequency ($\nu_{TA}^{\mathrm{max}} = \omega_{TA}^{\mathrm{max}}/(2\pi)$) in each material. The black oval regions in (a) and (c) indicate the weak AAA phase space due to AAA\#1 (BAs) and AAA\#2 (InP) selection rules in the frequency windows opened up by vanishing AAO and AOO scattering rates from the AAO and AOO\#1 selection rules in both materials. The red square regions in (a) and (b) show the weak AOO scattering rates from the AOO\#2 selection rule in BAs and AlSb respectively. The spectral $\kappa^{(3)}_{\mathrm{Pure}}$ for AlSb (h) also shows an anomalously large contribution from the optic phonons, which is completely suppressed by four-phonon scattering.}	\label{Fig3_HighKappaMat_PhononSpecificPlots_v4_InkscapeEdited}
\end{figure}

The seven compounds - BAs, BSb, BP, BN, SiC, GaN, and AlN have acoustic branches that become bunched together, thereby activating the AAA\#1 selection rule.  The signature of this activation is the sharp dip in the AAA scattering phase space (see Figs.~\ref{Fig3_HighKappaMat_PhononSpecificPlots_v4_InkscapeEdited} (a), S6, S7 (a), S8 (a)). BAs and BSb have been discussed in depth in the context of ultra-high $\kappa$ where the AAA\#1, AAO and AOO\#1 selection rules conspire to make three-phonon scattering unusually weak~\cite{lindsay_first-principles_2013_ref20}. The large pnictide to B mass ratios of BAs and BSb ($\sim$7 in BAs, $\sim$11 in BSb) give large A-O gaps that completely freeze out AAO scattering, and the narrow optic phonon bandwidths constrain AOO scattering of the acoustic phonons to low frequencies (see Figs. S1 (c) and (d) for phonon dispersions, Fig.~\ref{Fig3_HighKappaMat_PhononSpecificPlots_v4_InkscapeEdited} (a) and Fig. S6 (c) for the three-phonon phase space for BAs and BSb respectively). As seen in these figures, these two selection rules expose a large frequency window of only AAA scattering processes.  This combined with the activation of the AAA\#1 selection rule gives a small phase space for the lowest-order processes and results in significantly weaker three-phonon scattering rates and large contributions to $\kappa^{(3)}_{\mathrm{Pure}}$ (see Figs.~\ref{Fig3_HighKappaMat_PhononSpecificPlots_v4_InkscapeEdited}(a),~\ref{Fig3_HighKappaMat_PhononSpecificPlots_v4_InkscapeEdited}(d),~\ref{Fig3_HighKappaMat_PhononSpecificPlots_v4_InkscapeEdited}(g) for the results for BAs, and Figs. S6 (c), S13 (c) and S16 (c) for those of BSb). Upon including the four-phonon scattering the RT $\kappa^{(3)}_{\mathrm{Pure}}$ of BAs is sharply reduced from 3100 Wm$^{-1}$K$^{-1}$ to around 1400 Wm$^{-1}$K$^{-1}$.  In spite of this large reduction, we will show below that the four-phonon scattering in BAs is, in fact, much weaker for the acoustic phonons in the frequency range where the AAA three-phonon scattering rates are small, than in many other materials. We note that our calculations of $\kappa$ including three-phonon, four-phonon and phonon-isotope scattering are in good agreement with the measured data~\cite{tian_unusual_2018_ref19, kang_experimental_2018_ref27, li_high_2018_ref28} (Fig.~\ref{Fig2_HighKappaMat_kvsT_WithExpt} (a)).\\

\onecolumngrid

\twocolumngrid
 
The much smaller heavy-to-light atom mass ratios in BN, SiC, BP and AlN result in smaller A-O gaps in these materials.  As a result, AAO scattering dominates in the same frequency region as the dips in the AAA scattering rates. The implications of this \textsl{masking} of weak AAA scattering are discussed below.  The large $\kappa^{(3)}_{\mathrm{Pure}}$ in these materials occurs for the conventional reasons:  light atoms and stiff chemical bonds.  Measured values~\cite{chen_ultrahigh_2020, katre_exceptionally_2017_ref44, kang_thermal_2017_ref46, zheng_high_2018_ref47, morelli_IPCS_1993_ref45} are close to those predicted by the lowest-order theory. GaN has a relatively large Ga to N mass ratio of 5, which gives a large enough A-O gap to partially expose the dip in the AAA scattering rates (Fig. S8(a)). This contributes to a relatively large RT $\kappa^{(3)}_{\mathrm{Pure}} \sim$ 400 Wm$^{-1}$K$^{-1}$, consistent with prior calculations~\cite{lindsay_ab_2013_ref43}, but far lower than the RT $\kappa^{(3)}_{\mathrm{Pure}} \sim$ 3000 Wm$^{-1}$K$^{-1}$ obtained for BAs.\\

In BP, BN, SiC, GaN, InN and AlN, inclusion of four-phonon scattering has only a weak effect on $\kappa$ (3\%, 5\%, 9\%, 4\%, 13\%, and 15\% reductions in $\kappa^{(3)}_{\mathrm{Pure}}$ at RT for SiC, BP, BN, AlN, InN, and GaN, respectively) as seen in Fig. S5 and in Ref.~\cite{chen_ultrahigh_2020}. This is explained in part by the remarkably weak four-phonon scattering rates, which are found to be characteristic of the compounds with B, C and N atoms, as discussed in detail below.\\

\subsection{Materials influenced by AAA\#2 selection rule: InP, InAs, InSb, GaSb}
\begin{figure}[!ht]
\begin{center}
\includegraphics*[scale=0.45]{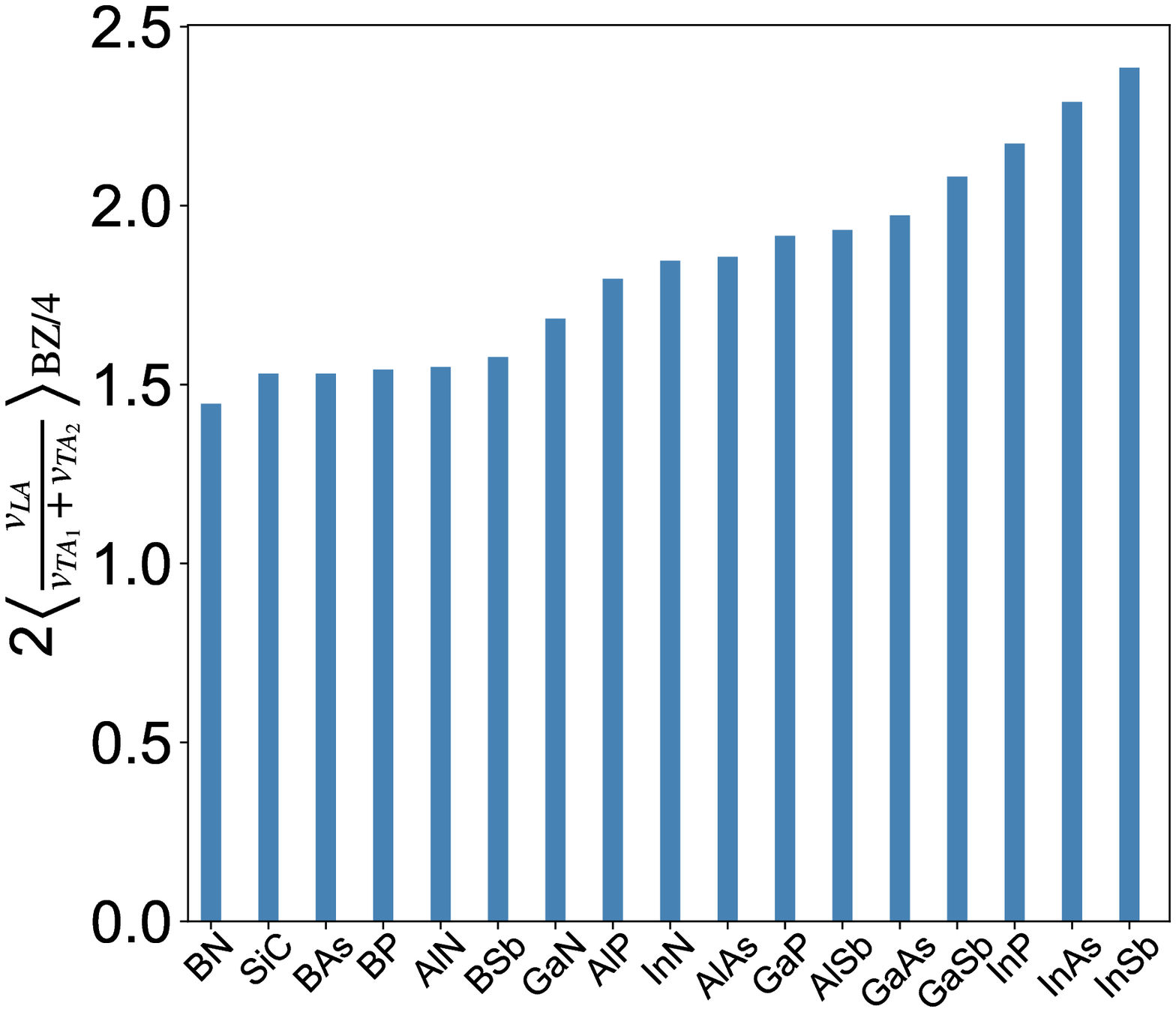}
\end{center}
\caption{Ratios of the LA phonon velocities to the mean TA phonon velocities averaged over $1/4^{\mathrm{th}}$ of the BZ centered at the $\Gamma$-point for seventeen zinc blende compounds. The compounds are listed in the ascending order of the mean LA-TA velocity ratio. Compounds with the largest velocity ratios have the AAA\#2 selection rule activated.}	\label{VelocityRatio_16Mat_BZby4}
\end{figure}

\begin{figure}[h]
\begin{center}
\includegraphics*[scale=0.4]{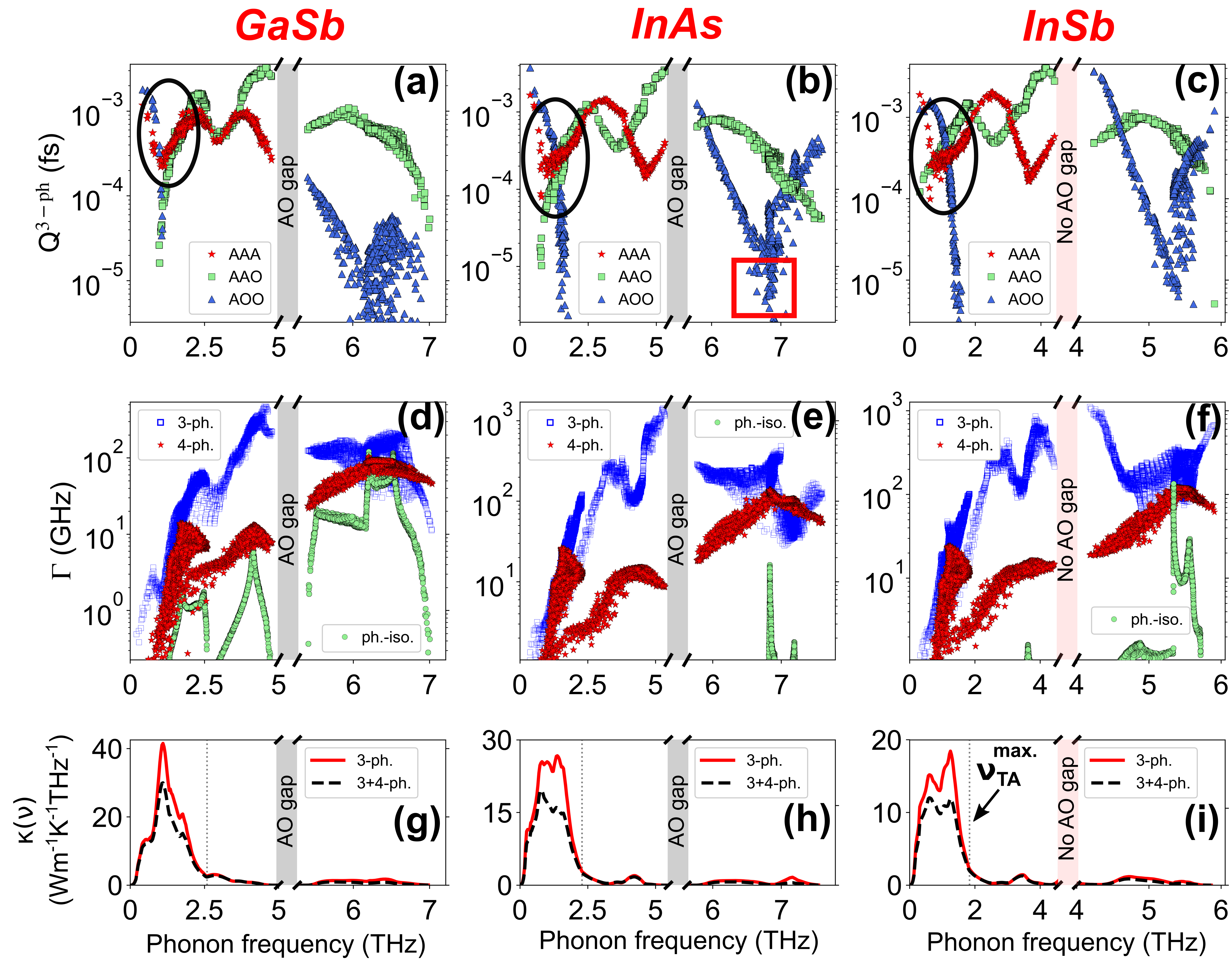}
\end{center}
\caption{Calculated three-phonon scattering phase space for different scattering channels (a)-(c), three-phonon, four-phonon and phonon-isotope scattering rates at 300 K (d)-(f) and spectral contributions to $\kappa^{(3)}_{\mathrm{Pure}}$ (solid red) and $\kappa^{(3+4)}_{\mathrm{Pure}}$ (dashed black) at 300 K (g)-(i), for GaSb (first column: (a), (d) and (g)), InAs (second column: (b), (e) and (h)) and InSb (third column (c), (f) and (i)). The dotted vertical lines in (g)-(i) show the location of the maximum TA phonon frequency ($\nu_{TA}^{\mathrm{max}} = \omega_{TA}^{\mathrm{max}}/(2\pi)$) in each material. The black oval regions in (a)-(c) indicate weak AAA phase space due to the AAA\#2 selection rule. However, unlike in Fig.~\ref{Fig3_HighKappaMat_PhononSpecificPlots_v4_InkscapeEdited} (a)-(c), relatively strong AAO and AOO scattering channels partially overlap with the weak AAA scattering frequency range, thus increasing the total three-phonon scattering rates in (d)-(f). The red square region in (b) shows the weak AOO scattering rates from the AOO\#2 selection rule, where, once again, AAO scattering channel is strong due to the small A-O gap in InAs.}	\label{Fig5_MediumKappaMat_PhononSpecificPlots_v2_InkscapeEdited}
\end{figure}

Fig.~\ref{VelocityRatio_16Mat_BZby4} ranks the ratio of LA to TA phonon group velocities, averaged over the center quarter of the BZ around the $\Gamma$-point for all seventeen compounds considered in this work.  Note that the choice of 1/4$^{\mathrm{th}}$ of the BZ is somewhat arbitrary and the same ordering would be obtained choosing e.g. 1/5$^{\mathrm{th}}$ of the BZ.  The four compounds - InP, InAs, InSb, and GaSb show the largest differences between the averaged LA and TA phonon group velocities, so more strongly activate the AAA\#2 selection rule. The most pronounced effect occurs in InP.  The large In to P mass ratio of 3.7 in InP creates a large A-O gap, and the optic phonon bandwidth is small (see Fig. S4(b)). Thus, AAO and AOO\#1 selection rules create a frequency window (1.5-3.5 THz) where only AAA scattering processes occur (red stars in Fig. 5 (c)). There, a significantly reduced phase space for AAA scattering results in the low frequency region (black oval in Fig. 5(c)). The striking effects of this restriction are seen in the spectral contributions to $\kappa^{(3)}_{\mathrm{Pure}}$ at RT shown in Fig.~\ref{Fig5_MediumKappaMat_PhononSpecificPlots_v2_InkscapeEdited} (i) where a large peak is seen around 1.5-2 THz.\\
 
In the 1.5-2 THz frequency region of weak AAA scattering, the RT four-phonon scattering rates are found to be comparable to or even exceed those of the three-phonon processes (Fig.~\ref{Fig5_MediumKappaMat_PhononSpecificPlots_v2_InkscapeEdited} (f)). This results in a large suppression in  by around 45\% at 300 K, which increases to almost 70\% at 750 K. These remarkably large reductions are supported by the measured  data for InP~\cite{kudman_thermal_1964_ref32, aliev1965thermal_ref31} as shown in Fig.~\ref{Fig4_MediumKappaMat_kvsT_WithExpt}(c), which is in excellent agreement with the results including four-phonon scattering over a wide temperature range.\\

Large differences in the LA and TA velocities in InAs, InSb and GaSb (Fig.~\ref{VelocityRatio_16Mat_BZby4}) also activate the AAA\#2 selection rule, which gives the low frequency dips in the AAA phase space identified by the black ovals in Figs.~\ref{Fig5_MediumKappaMat_PhononSpecificPlots_v2_InkscapeEdited}(a)-(c). However, the smaller heavy-to-light atom mass ratios compared to InP result in smaller frequency windows, only in a small portion of the frequency range where the dips in the AAA phase space occur are exposed. Since the dominant contributions to $\kappa$ from the lowest-order theory occur in this frequency region, where four-phonon scattering is also strong (see Fig.~\ref{Fig5_MediumKappaMat_PhononSpecificPlots_v2_InkscapeEdited}), a significant suppression in $\kappa^{(3)}_{\mathrm{Pure}}$ results. Four-phonon scattering suppresses the $\kappa^{(3)}_{\mathrm{Pure}}$ for each compound in this group, ranging from around 25-30\% at RT up to around 40-50\% at 750 K.

\onecolumngrid

\twocolumngrid

\subsection{Materials influenced by AOO\#2 selection rule: BAs, AlSb, AlAs, AlP, BP, GaAs, GaSb, GaP, InAs, InSb}
Optic phonon modes have only two possible channels for the lowest-order anharmonic decay: AAO and AOO.  The optic phonons in the ten listed compounds have regions of near degeneracy and so are strongly influenced by the AOO\#2 selection rule, as seen from the sharp dips in the phase space for AOO scattering (see Figs.~\ref{Fig3_HighKappaMat_PhononSpecificPlots_v4_InkscapeEdited} (a)-(b), ~\ref{Fig5_MediumKappaMat_PhononSpecificPlots_v2_InkscapeEdited} (a)-(c), S6 (b), S7 (b)-(c), S8 (b)-(c)). However, the relatively small heavy-to-light atom mass ratios in AlAs, AlP, BP, GaAs, GaSb, GaP, InAs, and InSb give no frequency window to expose the weak AOO scattering rates. Thus, the effects of the AOO\#2 selection rule on optic phonon lifetimes are masked in these compounds. The most extreme effects occur in BAs and AlSb because the large A-O gaps create frequency windows where AAO scattering is frozen out, thus exposing the regions of weak AOO scattering. The unusually weak ionicity in BAs~\cite{wentzcovitch_theoretical_1987_ref41} results in a near degeneracy of the three optic branches near the $\Gamma$-point of the BZ (see SM Fig. S1(c)). Then, the actions of AAO and AOO\#2 selection rules result in exceptionally small phase space for AOO scattering for the longitudinal optic (LO) and transverse optic (TO) phonons at the $\Gamma$-point (Fig.~~\ref{Fig3_HighKappaMat_PhononSpecificPlots_v4_InkscapeEdited}(a), red box). As a result, large room temperature contributions to $\kappa$ of around 130 Wm$^{-1}$K$^{-1}$ are predicted from the lowest-order theory (Fig.~\ref{Fig3_HighKappaMat_PhononSpecificPlots_v4_InkscapeEdited}(g)), significantly exceeding the total $\kappa$ of many common semiconductors. In contrast, BSb has a larger separation between the LO and TO phonons throughout the BZ (e.g., see SM Fig. S1(d)); thus the AOO\#2 selection rule has little influence on the optic phonons, which contribute negligibly to $\kappa$.\\

As in BAs, the large A-O gap in AlSb (Fig. S2 (d)) resulting from the large Sb to Al mass ratio creates a frequency window with no AAO scattering and exposes the weak AOO scattering phase space from the AOO\#2 selection rule (Fig.~\ref{Fig3_HighKappaMat_PhononSpecificPlots_v4_InkscapeEdited}(b), red box). However, in stark contrast to BAs, this occurs near the degeneracy of LO and TO$_2$ optic phonon modes in regions of the BZ far from the zone center and away from high symmetry directions (see SM Fig. S10 (d)). In these regions, TO$_2$ + A $\leftrightarrow$ LO processes become vanishingly small, while LO $\leftrightarrow$ TO$_1$ + A and TO$_2$ $\leftrightarrow$ TO$_1$ + A scattering is weak. The resulting small lowest-order optic phonon scattering rates (Fig.~\ref{Fig3_HighKappaMat_PhononSpecificPlots_v4_InkscapeEdited}(e)) give anomalously large contributions to $\kappa^{(3)}_{\mathrm{Pure}}$ of around 90 Wm$^{-1}$K$^{-1}$ (Fig.~\ref{Fig3_HighKappaMat_PhononSpecificPlots_v4_InkscapeEdited}(h)).\\

Four-phonon scattering in BAs and AlSb completely suppress the optic phonon lifetimes (Figs.~\ref{Fig3_HighKappaMat_PhononSpecificPlots_v4_InkscapeEdited} (d), (e), (g) and (h)), in striking contrast to predictions from the lowest-order theory. Fig.~\ref{Fig3_HighKappaMat_PhononSpecificPlots_v4_InkscapeEdited}(e) shows that four-phonon scattering rates for the optic phonons are much larger than their three-phonon counterparts at 300 K for most of the optic phonon bandwidth. Surprisingly, this is true even at a low temperature of 100 K, where the influence of four-phonon scattering is expected to be weak even for strongly anharmonic materials~\cite{ravichandran_unified_2018_ref7}. The AOAO and OOOO four-phonon processes dominate the scattering as shown in the (SM section S9, Figs. S23(c) and S24 (d)). For AlSb, optic phonon contributions dominate the calculated RT $\kappa^{(3)}_{\mathrm{Pure}}$, so including four-phonon scattering gives substantially reduced $\kappa^{(3+4)}_{\mathrm{Pure}}$ indicating that the lowest-order theory is failing drastically. In fact, our results indicate that there is very little hope for identifying materials with large contributions to $\kappa$ coming from the optic phonons, since (a) at moderate-to-high temperatures, four-phonon scattering dominates any frequency region of weak three-phonon scattering for the optic phonons driven by the AAO and AOO\#2 selection rules, thereby significantly lowering while (b) at low temperatures, where four-phonon scattering rates can get weaker than their three-phonon counterparts, the heat capacity of the optic phonons becomes vanishingly small owing to their high frequency scale (see SM section S7).\\

An interesting additional failure of the lowest-order theory for AlSb is its prediction of a large isotope effect (enhancement in $\kappa$ upon isotopic enrichment), which is seen in the difference between the dashed black and dashed red curves in Fig.~\ref{Fig2_HighKappaMat_kvsT_WithExpt}(b). This is quantified by $P = 100\left(\kappa_{\mathrm{Pure}}/\kappa_{\mathrm{Nat.}} - 1\right)$, which remains above 50\% between 100 K and 300 K ignoring four-phonon interactions. In compounds where the heavy-to-light ratio of the masses of the two constituent atoms is large, phonon-isotope scattering is weak throughout the BZ if the vibrating atom is isotopically pure~\cite{lindsay_first-principles_2013_ref20, lindsay_phonon-isotope_2013_ref42}. This is the case for AlSb ($m_{\mathrm{Sb}}/m_{\mathrm{Al}}$ = 4.5) where the light, isotopically pure Al atoms dominate the vibrational component of the optic modes while the heavy Sb atoms, which have a large isotope mix (57\% $^{121}$Sb, 43\% $^{123}$Sb), hardly move. The resulting weak phonon-isotope scattering shows a striking contrast to the opposite behavior in BAs, where the heavy atom (As) is isotopically pure but the light atom (B) is not (compare phonon-isotope scattering rates in Figs.~\ref{Fig3_HighKappaMat_PhononSpecificPlots_v4_InkscapeEdited}(d) and (e) ). Nevertheless, this weak phonon-isotope scattering strongly suppresses $\kappa^{(3)}_{\mathrm{Pure}}$ in AlSb because of the already weak AOO scattering. When four-phonon interactions are included, the $P$ of AlSb drops to between 10\% and 1\% in the same temperature range.  Note that the calculated $\kappa^{(3+4)}_{\mathrm{Nat.}}$ is in much better agreement with the measured $\kappa$ data in Fig.~\ref{Fig2_HighKappaMat_kvsT_WithExpt}(b) than $\kappa^{(3)}_{\mathrm{Nat.}}$.\\

We note that large suppression of anharmonic optic phonon lifetimes and of thermal conductivity by four-phonon scattering has been predicted in prior work for BAs~\cite{feng_four-phonon_2017_ref6} and AlSb~\cite{yang_stronger_2019}.  These works attributed the large lowest-order optic phonon lifetimes to freezing out of AAO scattering by the large A-O gaps.  While it is true that AAO scattering is limited, this does not explain why AOO scattering is weak in BAs and AlSb.  As discussed above BSb also has a large A-O gap that removes AAO processes but has much smaller contributions to $\kappa^{(3)}_{\mathrm{Pure}}$ than in either BAs or AlSb because its larger LO-TO splitting results in larger minimum AOO scattering rates. The new AOO selection rule identified in the present work provides the missing explanation for why AOO scattering becomes weak in both BAs and AlSb but not in BSb.\\

\subsection{Other compounds: GaP, AlAs, GaAs, AlP}
The heavy-to-light atom mass ratios in GaP (2.3) and AlAs (2.6) are large enough for AAO and AOO\#1 selection rules to create frequency gaps where only AAA scattering occurs.  Even though no Type 2 selection rules are activated in these regions, the lack of AAO and AOO scattering in these frequency windows give relatively weak three-phonon scattering, and correspondingly large contributions to $\kappa^{(3)}$ result. Four-phonon scattering suppresses $\kappa^{(3)}$ by around 25\% at 300K and around 45\% at 750K.\\

Relatively small reductions are found in the calculated values of $\kappa^{(3)}$ in AlP and GaAs upon inclusion of four-phonon scattering - around 10-15\% at RT. The constituent elements of the two compounds have similar masses resulting in small A-O gaps, so AAO scattering extends throughout the range of the acoustic and the optic phonons (Figs. S8(c) for GaAs, S7(b) for AlP). Neither of the AAA selection rules are activated by features in the phonon dispersions for the two materials.\\
 
The results for AlP and GaAs highlight the fact that anomalous suppression of $\kappa^{(3)}$ by four-phonon scattering is unlikely to occur for strongly-bonded compounds with similar masses since no frequency window is opened by the Type 1 selection rules.  A frequency window will also not occur in strongly polar materials having large LO-TO splitting and large optic phonon bandwidths.  An example of this behavior is found is MgO~\cite{ravichandran_non-monotonic_2019}, which has no frequency window opened by the selection rules. We have calculated the four-phonon scattering rates in MgO and find them to be comparable to those in the non-BCN zinc blende compounds. Since no selection rules are activated, three-phonon scattering is also strong.  A reduction in $\kappa^{(3)}$ of about 15\% at is found at RT, consistent with that found in GaAs and AlP. 

\section{Four-phonon scattering and its weakness in compounds with boron, carbon or nitrogen} 
Comparison of the four-phonon scattering rates for the seventeen materials studied shows several striking features. First, the four-phonon scattering rates generally increase with increasing frequency, being the strongest for the optic phonons.  In particular, they are far larger than the small lowest-order optic phonon scattering rates in BAs and AlSb responsible for the erroneously large contributions to $\kappa$.  As noted above, this finding suggests that it is not likely to be possible to get large contributions from optic phonons in any compound, and any findings to the contrary in the lowest-order phonon-phonon theory are likely incomplete and point simply to the need to consider four-phonon interactions to obtain accurate results.  Nevertheless, the dominance of four-phonon anharmonic decay rates for the optic phonons compared with their three-phonon counterparts that we find in BAs, AlSb, InP, AlAs and BSb around $\Gamma$ (see Figs.~\ref{Fig3_HighKappaMat_PhononSpecificPlots_v4_InkscapeEdited}(d)-(f), S14(c) and S13(c) respectively) suggests that these signatures should be observable in e.g. temperature dependent Raman linewidth measurements.\\
 
Second, the four-phonon scattering rates for the nine non-BCN compounds i.e. those that do not contain B, C, or N atoms (GaAs, GaP, GaSb, AlSb, AlP, AlAs, InP, InAs, InSb) are remarkably similar in magnitude.  An example of this is seen in Figs.~\ref{Fig3_HighKappaMat_PhononSpecificPlots_v4_InkscapeEdited} (e) and (f) for AlSb and InP, respectively.  Comparisons for the other compounds can be made in Figs.~\ref{Fig5_MediumKappaMat_PhononSpecificPlots_v2_InkscapeEdited} (d)-(f), and Figs. S14 (b)-(c) and S15 (b)-(c) in the SM.  These similarities across materials are in striking contrast to the analogous lowest-order scattering rates.  This highlights a fundamental difference between the lowest-order phonon-phonon interactions and those of higher order.  The lowest-order processes are constrained by selection rules; features in the phonon dispersions that activate these selection rules can significantly reduce the lowest-order phonon-phonon scattering rates. In contrast, the phase space for four-phonon scattering shows remarkably little structure across materials, being immune to shifts in the A-O gap, or to the \textsl{bunching together} of the acoustic or optic branches.  The four-phonon scattering rates of the acoustic phonons for all compounds are dominated by AOAO and AAAA processes, and for those, fourth-order bond anharmonicity and phonon frequency scales act in opposition, maintaining relative uniformity in the scattering strengths.  This is discussed in more detail in the SM section S10.  Based on the relatively uniform strength of the four-phonon scattering rates and lack of selection rules influencing them, it seems reasonable to expect that higher-order phonon-phonon interactions beyond four-phonon scattering (five-phonon and beyond) should play a negligible role in the thermal conductivity of this group of materials.\\
 
The most remarkable feature is that the four-phonon scattering rates for the BCN compounds, i.e. those containing B, C or N atoms (BAs, BSb, BP, BN, SiC, GaN, AlN and InN) are much smaller than those for the non-BCN compounds.  This is clearly seen by comparing Figs.~\ref{Fig3_HighKappaMat_PhononSpecificPlots_v4_InkscapeEdited}(d)-(f), which show that the four-phonon scattering rates in BAs are an order of magnitude smaller than those for InP and AlSb.  Figs. S13, S14 and S15, and Fig. 3B in Ref.~\cite{chen_ultrahigh_2020} show that the four-phonon scattering rates for BSb, BP, BN, SiC, GaN, AlN and InN are similarly small compared to their non-BCN counterparts. The four-phonon scattering rates at 750 K show the same trends as those identified at RT.  We have traced this behavior to fundamental bonding differences between BCN and non-BCN compounds.  BCN compounds have stronger bonding, reflected in their smaller lattice constants and higher phonon frequency scales compared to the non-BCN compounds.  However, the fourth-order anharmonic component of the bonds in the BCN compounds tends to be weaker relative to the harmonic component, compared with those in the non-BCN compounds. The implications of this finding are discussed below. Interestingly, recent calculations of thermal transport in ZrC~\cite{mellan_electron_2019}, a metal with rock salt structure, also show unusually weak four-phonon scattering rates and correspondingly small reductions in the phonon-phonon limited $\kappa$, consistent with our findings for the zinc blende BCN compounds.\\
  
Exploiting our finding that the AOAO processes are dominant in all the considered materials, we have developed a simple model that captures the weaker four-phonon scattering strength in the BCN compounds compared with the non-BCN materials.  This model is discussed in the SM section 10, and findings are plotted in Fig. S27.
 
\section{Discussion}
From the above findings we make the following additional observations:\\

\textsl{Significance of weak four-phonon scattering in BCN compounds}: While the $\kappa$ of BAs is significantly suppressed by four-phonon scattering, it still achieves an ultra-high RT value of $\kappa^{(3+4)}_{\mathrm{Nat.}}$ of around 1300 Wm$^{-1}$K$^{-1}$, confirmed by measurements~\cite{tian_unusual_2018_ref19, li_high_2018_ref28, kang_experimental_2018_ref27}, about two to three times larger than other high $\kappa$ compounds such as SiC, BP, copper and silver.  To test the importance of its unusually weak four-phonon processes, we artificially increased the RT four-phonon scattering rates in BAs by a factor of ten, making them comparable to those in the non-BCN compounds. The resulting RT $\kappa$ of BAs is reduced by over 70\% to 330 Wm$^{-1}$K$^{-1}$. Similarly increased four-phonon scattering rates in GaN and BP yield 50\% and 30\% decreases in the RT $\kappa_{\mathrm{Pure}}$ values for these compounds. This highlights one of the fortuitous benefits that nature gives to the BCN compounds, without which their $\kappa$'s could be much lower.\\

\textsl{High $\kappa$ in non-BCN compounds unlikely}: The larger four-phonon scattering found in the non-BCN materials suggests that finding a material with high $\kappa$ among them will be challenging.  Indeed, note that of the eight compounds above, showing small deviations from the predictions of the lowest-order theory upon inclusion of four-phonon scattering, only two (GaAs and AlP) are non-BCN compounds.  Conventional guidelines for achieving high $\kappa$, as described by Slack~\cite{slack_nonmetallic_1973_ref48}, require the combination of light atoms and exceptionally strong bonding, with diamond being the prototype.  These criteria already rule out the non-BCN compounds. The alternative paradigm exemplified by BAs~\cite{lindsay_first-principles_2013_ref20} in principle allows for the possibility of a high $\kappa$ material composed of non-BCN constituent atoms.  Such a material would require phonon dispersions for which the selection rules give small three-phonon scattering phase space in a frequency region where the four-phonon scattering is weak.  This does not occur in any of the studied non-BCN compounds, where instead, frequency windows with small total three-phonon phase space coincide with strong four-phonon scattering regions.\\
 
\textsl{Masking of weak AAA and AOO scattering}:  The four compounds, BAs, BSb, AlSb and InP, showing the catastrophic failures of the lowest-order theory have the largest A-O gaps, which remove the AAO processes. This feature is clearly a critical requirement to achieve frequency windows in which anomalously large phonon lifetimes are possible.  To highlight the importance of this requirement, consider BP and SiC compared with BAs.  The AAA, AOO and four-phonon scattering rates in SiC, BP and BAs are similar.  Thus, the reason SiC and BP have lower $\kappa$ than BAs stems from the AAO scattering rates, which are much larger than the AAA scattering rates in the frequency range of the AAA minima.  To illustrate how important these are, we have also calculated the $\kappa$ of SiC and BP with AAO processes artificially removed.  We find that the RT $\kappa^{(3)}_{\mathrm{Pure}}$ of BP in this case is over 6000 Wm$^{-1}$K$^{-1}$, while including four-phonon scattering gives $\kappa^{(3+4)}_{\mathrm{Pure}}$ of around 2600 Wm$^{-1}$K$^{-1}$, roughly twice the corresponding values in BAs.  For SiC the corresponding values are 3800 Wm$^{-1}$K$^{-1}$ and 2700 Wm$^{-1}$K$^{-1}$.  While it is unfortunately not possible to remove these processes in reality, they could in principle be tuned.  For example, the application of hydrostatic pressure can preferentially shift the optic phonons to higher frequencies relative to the acoustic phonons.  This would weaken AAO processes and possibly give unusually rapid increase of $\kappa$ in SiC and BP.\\

Similar behavior is found for the optic phonons.  The compounds, AlAs, AlP, BP, GaAs, GaSb, GaP, InAs and InSb have regions of weaker AOO scattering rates due to the AOO\#2 selection rule than does AlSb. Thus, even larger contributions from optic phonons would be predicted from the lowest order-theory for these compounds if not for their comparably small A-O gaps, which give strong AAO scattering where the AOO optic phonon scattering is small.  As a result, optic phonons in these compounds contribute negligibly to $\kappa$ even in the lowest-order theory.\\

\textsl{Complementary features activating AAA\#1 and AAA\#2 selection rules:}  Figure~\ref{VelocityRatio_16Mat_BZby4} shows that those compounds that more strongly activate AAA\#1 or AAA\#2 selection rules lie on opposite ends of the velocity ratio plot. Thus, compounds that have acoustic branches that are bunched together away from the center of the BZ (BAs, BSb, BP, BN, SiC, AlN, GaN) and so activate the AAA\#1 selection rule, generally have smaller differences between near-zone center LA and TA velocities.  In contrast, compounds with large differences in their near-zone center LA and TA velocities (e.g. InSb, InAs, InP, GaSb), and so activate the AAA\#2 selection rule, do not show the acoustic branch bunching behavior or do not have it exposed by the Type 1 selection rules.\\

\textsl{Importance of new Type 2 selection rules}: Prior studies have already identified the importance of the AAA\#1 Type 2 selection rule in explaining the anomalously weak three-phonon phase space in BAs and BSb. Yet, this selection rule cannot explain the unusually large optic phonon contributions to $\kappa^{(3)}_{\mathrm{Pure}}$ in BAs and AlSb, nor can it explain the large acoustic phonon contributions to $\kappa^{(3)}_{\mathrm{Pure}}$ in InP, InAs, InSb and Gasb.  The two new Type 2 selection rules, AAA\#2 and AOO\#2, are required for full understanding of the behavior in the studied ZB materials.\\
  
\textsl{Restricted spectral contributions to thermal conductivity}:  In all of the studied compounds, we find that the dominant contributions to $\kappa$ occur below the highest TA phonon frequency, $\nu^{\mathrm{max}}_{\mathrm{TA}}$.  Above $\nu^{\mathrm{max}}_{\mathrm{TA}}$, LA phonons exhibit strong decay via AAA processes as well as participating in AAO processes resulting in small contributions to $\kappa$ in this frequency region.  The $\nu^{\mathrm{max}}_{\mathrm{TA}}$ are marked in the spectral $\kappa$-contribution figures for each material (Figs.~\ref{Fig3_HighKappaMat_PhononSpecificPlots_v4_InkscapeEdited}(d)-(f),~\ref{Fig5_MediumKappaMat_PhononSpecificPlots_v2_InkscapeEdited} (d)-(f) and S16-S18) to illustrate this behavior.\\

\textsl{Distinction between phase space and matrix element selection rules}: The selection rules discussed in the present work are connected to the phase space for three-phonon scattering.  They depend on specific features being present in phonon dispersions, as described in Section II.  It is important to distinguish these from a different kind of selection rule, which can cause the matrix elements for phonon-phonon scattering to vanish.  Such selection rules have been identified in systems of reduced dimensionality and are connected to the underlying symmetry in some crystals~\cite{lindsay_lattice_2009, lindsay_flexural_2010, pandey_symmetry-driven_2018, cammarata_phononphonon_2019}.  Their incorporation can improve efficiency in schemes for computing phonon-phonon processes, as recently discussed~\cite{cammarata_phononphonon_2019}.\\ 

\textsl{Effect of impurities on the measured $\kappa$}:  The large differences seen between the $\kappa$ values calculated only using the lowest-order theory and the measured data for many of the studied materials cannot be explained by possible impurities in the measured samples.  The presence of impurities gives a weaker temperature dependence to $\kappa$ compared to the predictions from the lowest-order theory, since phonon-impurity scattering is temperature independent, as described previously~\cite{tian_unusual_2018_ref19}.  In contrast, the measured $\kappa\left(T\right)$ shows a stronger temperature dependence than seen in many of the lowest-order calculations. The calculations including both three-phonon and four-phonon scattering processes also gives stronger T-dependence~\cite{feng_four-phonon_2017_ref6, tian_unusual_2018_ref19} and are both qualitatively and quantitatively in agreement with the experimentally measured temperature-dependent $\kappa$.\\

\textsl{Impact of selection rules in other materials}:  The selection rules on three-phonon scattering are completely general and are not restricted to ZB compounds. Creation of frequency windows in Fig.~\ref{Fig2_Schematic_v3} benefits from materials that have low ionicity (e.g., several ZB compounds in this study), which gives relatively small optic phonon bandwidths, and large heavy-to-light mass ratios. For materials with more than two atoms in the unit cell, the additional scattering channels resulting from the increased number of optic phonon branches will make the needed frequency windows harder to achieve for acoustic phonons, so the selection rules are less likely to affect the acoustic phonons in such compounds.  But, large frequency gaps between bands of optic phonons can be found in many compounds~\cite{togolink} so erroneously large anharmonic phonon decay rates and contributions to $\kappa$ from some optic phonons could occur from activation of selection rules such as AOO\#2 and OOO.  For such compounds, the selection rule picture can provide useful guidance for understanding measurements giving lower $\kappa$ and stronger temperature dependence of anharmonic phonon decay than predicted from calculations including only three-phonon scattering. In crystals with one atom in the unit cell, only AAA processes occur. We are not aware of any such materials that are semiconductors or insulators, and in metals, $\kappa$ is typically dominated by the electronic contribution. For the lattice contribution, inclusion of electron-phonon interactions could mask any influence of the selection rules on the lowest-order phonon-phonon interactions.
 
\section{Summary}
\textsl{Ab initio} theories of the anharmonic properties of materials continue to rely on descriptions that include only the lowest-order phonon-phonon interactions. The vastly different results obtained here for phonon lifetimes and thermal conductivity with and without four-phonon interactions for the many considered materials contradict prevailing understanding of anharmonic phonon decay and thermal conduction in weakly anharmonic crystals, where four-phonon interactions have frequently been assumed to be unimportant.\\

We have identified new selection rules on three-phonon scattering dictated entirely by features in the phonon dispersions and have shown that the full set of selection rules listed in Section II and the trends found in four-phonon scattering strengths provide a useful framework to understand the anharmonic phonon decay and thermal conductivity of weakly anharmonic insulating crystals. In particular, the new selection rules are critical in explaining the failure of the lowest-order theory in describing thermal transport in many zinc blende compounds.\\

In compounds that do not contain boron, carbon or nitrogen, four-phonon scattering suppresses the erroneously large phonon lifetimes and thermal conductivities predicted by the lowest-order theory.  Conversely, in compounds that do contain boron, carbon, and nitrogen atoms, large phonon lifetimes and thermal conductivities predicted in the lowest-order theory are less affected, because of the exceptionally weak four-phonon interactions.  These findings help explain the high thermal conductivities found in the technologically important compounds such as BAs, BP, and SiC, and provide critical guidance in the search for new materials with high thermal conductivity.\\

Inclusion of four-phonon interactions in \textsl{ab initio} anharmonic decay rates and transport calculations is computationally challenging, and such calculations are currently performed by only a small number of groups~\cite{feng_quantum_2016_ref5, feng_four-phonon_2017_ref6, ravichandran_unified_2018_ref7, xia_revisiting_2018_ref8, xia_anharmonic_2018_ref9, chen_ultrahigh_2020, ravichandran_non-monotonic_2019, yang_stronger_2019, mellan_electron_2019, tian_unusual_2018_ref19}.  Recent work has shown that for strongly anharmonic materials, it is essential to include four-phonon processes~\cite{ravichandran_unified_2018_ref7, xia_revisiting_2018_ref8, xia_anharmonic_2018_ref9}.  The present study shows that, even in weakly anharmonic materials that do not contain B, C or N atoms, much stronger four-phonon scattering occurs, and points to the possible need to include four-phonon scattering in future thermal transport calculations even in such strongly bonded compounds.  While this suggestion has been made before~\cite{feng_four-phonon_2017_ref6}, the large number of compounds studied in the present work and comparably larger four-phonon scattering strengths found in the non-BCN compounds adds strong quantitative and statistical support to this proposition.  Further calculations will be required to fully establish this point.

\section{Methods}
For all materials, temperature dependent lattice parameters are obtained by minimizing the Helmholtz free energy $F=\Phi_0 + F_H + F_A$, where $\Phi_0$ is the energy of the lattice atoms in their equilibrium positions, and $F_H$ is the harmonic part and $F_A$ is the anharmonic part, which includes both third and fourth order terms~\cite{ravichandran_unified_2018_ref7}. To obtain phonon modes, the second-order interatomic force constants (IFCs) are calculated using density functional perturbation theory (DFPT) as implemented in the QUANTUM ESPRESSO package~\cite{giannozzi_quantum_2009}.  All calculations in this work were performed using norm-conserving pseudopotentials with the local density approximation exchange correlation functional. Converged parameters for the density functional theory (DFT) calculations are provided in the SM section S14.\\

The third and fourth-order anharmonic IFCs required to obtain three-phonon and four-phonon scattering rates are calculated using a thermal stochastic snapshot technique.  The method has been described in detail in Ref.~\cite{ravichandran_unified_2018_ref7}. Temperature dependent anharmonic renormalization of phonon modes~\cite{ravichandran_unified_2018_ref7} was included but did not significantly affect phonon modes, scattering rates and thermal conductivities for the considered materials and temperature ranges.  To calculate the thermal conductivity of a material, the Boltzmann equation for phonon transport including three-phonon, four-phonon and phonon-isotope scattering was solved using an iterative approach for the non-equilibrium phonon distribution function created from an assumed small temperature gradient across the considered materials. For all materials and all temperatures, the phonon Boltzmann equation was solved on a 51X51X51 $\mathbf{q}$-grid.  Four-phonon scattering rates were obtained on 17X17X17 $\mathbf{q}$-grids and interpolated to the finer 51X51X51 $\mathbf{q}$-grids (see SM section S13 for details).  Full calculations (without interpolation) were also performed on 21X21X21 $\mathbf{q}$-grids to check the convergence of the four-phonon scattering rates (see SM section S12). The phonon Boltzmann equation and the expressions for the scattering rates are given in the Appendix.\\

\section{Acknowledgments}
This work was supported by the Office of Naval Research under a MURI, Grant No. N00014-16-1-2436.\\

N. K. R and D. B. originated the research. N.K.R. performed the \textsl{ab initio} calculations.  N.K.R. and D.B. analyzed the results and wrote the manuscript. Both authors studied, commented on and edited the manuscript.\\

The authors declare no competing financial interests.

\appendix*
\section{Phonon Boltzmann equation, scattering rates and thermal conductivity}
The phonon Boltzmann equation for an applied temperature gradient, $\Delta T$, is:
\begin{align}
\mathbf{v}_\lambda\cdot\nabla T\frac{\partial n^0_\lambda}{\partial T} = \frac{\partial n_\lambda}{\partial t}\Bigg|_{\mathrm{collisions}}	\label{PBE}
\end{align}
where $\mathbf{v}_\lambda$ is the phonon group velocity and $n^0_\lambda = 1/(e^{\hbar\omega_\lambda/(k_BT)} - 1)$ is the equilibrium Bose distribution at temperature, $T$, and for phonon mode $\lambda\sim\left(\mathbf{q}, j\right)$ and frequency  $\omega_\lambda$.\\

The non-equilibrium phonon distribution function, linearized in assumed small $\Delta T$ is:
\begin{align}
n_\lambda = n^{0}_\lambda + n^0_\lambda\left(n^0_\lambda + 1\right)\mathbf{F}_\lambda\cdot\left(-\nabla T\right)
\end{align}
The function, $\mathbf{F}_\lambda$, is obtained by solving the linearized phonon Boltzmann equation including three-phonon, four-phonon and phonon-isotope scattering:
\onecolumngrid

\begin{align}
\mathbf{F}_\lambda &= \mathbf{F}^0_\lambda + \tau^{\left(\mathrm{tot}\right)}_\lambda\Bigg\{\sum_{\lambda_1\lambda_2}\left[W^{\left(+\right)}_{\lambda\lambda_1\lambda_2}\left(\mathbf{F}_{\lambda_2} - \mathbf{F}_{\lambda_1}\right) + \frac{1}{2}W^{\left(-\right)}_{\lambda\lambda_1\lambda_2}\left(\mathbf{F}_{\lambda_2} + \mathbf{F}_{\lambda_1}\right)\right] +  \sum_{\lambda_1}W^{\mathrm{iso}}_{\lambda\lambda_1}\mathbf{F}_{\lambda_1}\nonumber\\
&\ \ \ \ \ \ \ \ \ \ \ \ \ \ \ \ \ + \sum_{\lambda_1\lambda_2\lambda_3}\Bigg[\frac{1}{6}Y^{\left(1\right)}_{\lambda\lambda_1\lambda_2\lambda_3}\left(\mathbf{F}_{\lambda_1} + \mathbf{F}_{\lambda_2} + \mathbf{F}_{\lambda_3}\right) + \frac{1}{2}Y^{\left(2\right)}_{\lambda\lambda_1\lambda_2\lambda_3}\left(\mathbf{F}_{\lambda_2} + \mathbf{F}_{\lambda_3} - \mathbf{F}_{\lambda_1}\right) \nonumber\\
&\ \ \ \ \ \ \ \ \ \ \ \ \ \ \ \ \ \ \ \ \ \ \ \ \ \ + \frac{1}{2}Y^{\left(3\right)}_{\lambda\lambda_1\lambda_2\lambda_3}\left(\mathbf{F}_{\lambda_3} - \mathbf{F}_{\lambda_2} - \mathbf{F}_{\lambda_1}\right)\Bigg]\Bigg\}	\label{PBE_4ph}
\end{align}
where,
\begin{align}
\mathbf{F}^0_\lambda = \hbar\omega_\lambda\mathbf{v}_\lambda\tau^{\left(\mathrm{tot}\right)}_\lambda/k_BT^2
\end{align}
\begin{align}
1/\tau_{\lambda}^{(\mathrm{tot})} = 1/\tau_{\lambda}^{(\mathrm{3-ph})} + 1/\tau_{\lambda}^{(\mathrm{4-ph})} + 1/\tau_{\lambda}^{(\mathrm{ph.-iso.})}
\end{align}
\begin{align}
1/\tau^{\left(\mathrm{3-ph.}\right)}_\lambda =  \sum_{\lambda_1\lambda_2}\left[W^{\left(+\right)}_{\lambda\lambda_1\lambda_2} + \frac{1}{2}W^{\left(-\right)}_{\lambda\lambda_1\lambda_2}\right]
\end{align}
\begin{align}
1/\tau_{\lambda}^{(\mathrm{ph.-iso.})} = \sum_{\lambda_1}W^{\mathrm{iso}}_{\lambda\lambda_1}
\end{align}
\begin{align}
1/\tau^{\left(\mathrm{4ph}\right)}_\lambda = \sum_{\lambda_1\lambda_2\lambda_3}\left[\frac{1}{6}Y^{\left(1\right)}_{\lambda\lambda_1\lambda_2\lambda_3} + \frac{1}{2}Y^{\left(2\right)}_{\lambda\lambda_1\lambda_2\lambda_3} + \frac{1}{2}Y^{\left(3\right)}_{\lambda\lambda_1\lambda_2\lambda_3}\right]
\end{align}

The three-phonon scattering probabilities are:
\begin{align}
W^{\left(+\right)}_{\lambda\lambda_1\lambda_2} &= \frac{2\pi}{\hbar ^2}\left|\Phi_{\lambda\lambda_1\left(-\lambda_2\right)}\right|^2\left(n^0_{\lambda_1} - n^0_{\lambda_2}\right)\delta\left(\omega_\lambda + \omega_{\lambda_1} - \omega_{\lambda_2}\right)\nonumber\\
W^{\left(-\right)}_{\lambda\lambda_1\lambda_2} &= \frac{2\pi}{\hbar ^2}\left|\Phi_{\lambda\left(-\lambda_1\right)\left(-\lambda_2\right)}\right|^2\left(1 + n^0_{\lambda_1} + n^0_{\lambda_2}\right)\delta\left(\omega_\lambda - \omega_{\lambda_1} - \omega_{\lambda_2}\right)
\end{align}
The four-phonon scattering probabilities are:
\begin{align}
Y^{\left(1\right)}_{\lambda\lambda_1\lambda_2\lambda_3} &= \frac{2\pi}{\hbar ^2}\left|\Phi_{\lambda\left(-\lambda_1\right)\left(-\lambda_2\right)\left(-\lambda_3\right)}\right|^2\frac{n^0_{\lambda_1}n^0_{\lambda_2}n^0_{\lambda_3}}{n^0_\lambda}\delta\left(\omega_\lambda - \omega_{\lambda_1} - \omega_{\lambda_2} - \omega_{\lambda_3}\right)\nonumber\\
Y^{\left(2\right)}_{\lambda\lambda_1\lambda_2\lambda_3} &= \frac{2\pi}{\hbar ^2}\left|\Phi_{\lambda\lambda_1\left(-\lambda_2\right)\left(-\lambda_3\right)}\right|^2\frac{\left(1 + n^0_{\lambda_1}\right)n^0_{\lambda_2}n^0_{\lambda_3}}{n^0_\lambda}\delta\left(\omega_\lambda + \omega_{\lambda_1} - \omega_{\lambda_2} - \omega_{\lambda_3}\right)\nonumber\\
Y^{\left(3\right)}_{\lambda\lambda_1\lambda_2\lambda_3} &= \frac{2\pi}{\hbar ^2}\left|\Phi_{\lambda\lambda_1\lambda_2\left(-\lambda_3\right)}\right|^2\frac{\left(1 + n^0_{\lambda_1}\right)\left(1 + n^0_{\lambda_2}\right)n^0_{\lambda_3}}{n^0_\lambda}\delta\left(\omega_\lambda + \omega_{\lambda_1} + \omega_{\lambda_2} - \omega_{\lambda_3}\right)
\end{align}
The three-phonon and four phonon matrix elements are
\begin{align}
\Phi_{\lambda\lambda_1\lambda_2} &= \Phi_{\mathbf{q}s, \mathbf{q}_1s_1, \mathbf{q}_2s_2}\nonumber\\
&=\left(\hbar/2\right)^{3/2}\left(1/N_0^{1/2}\right)\left[\omega_{\mathbf{q}s}\omega_{\mathbf{q}_1s_1}\omega_{\mathbf{q}_2s_2}\right]^{-1/2}\nonumber\\
&\ \ \ \ \times \sum_{NP}\sum_{\mu\nu\pi}\sum_{\alpha\beta\gamma}\Phi_{\alpha\beta\gamma}\left(0\mu, N\nu, P\pi\right)\left(M_\mu M_\nu M_\pi\right)^{-1/2}\nonumber\\
&\ \ \ \ \times e^{i\mathbf{q}_1\cdot\mathbf{R}\left(N\right)}e^{i\mathbf{q}_2\cdot\mathbf{R}\left(P\right)}\nonumber\\
&\ \ \ \ \times w_\alpha\left(\mathbf{q}s,\mu\right)w_\beta\left(\mathbf{q}_1s_1,\nu\right)w_\gamma\left(\mathbf{q}_2s_2,\pi\right)		\label{3_ph_mat}
\end{align}
and,
\begin{align}
\Phi_{\lambda\lambda_1\lambda_2\lambda_3} &= \Phi_{\mathbf{q}s, \mathbf{q}_1s_1, \mathbf{q}_2s_2, \mathbf{q}_3s_3}\nonumber\\
&= \left(\hbar/2\right)^2\left(1/N_0\right)\left[\omega_{\mathbf{q}s}\omega_{\mathbf{q}_1s_1}\omega_{\mathbf{q}_2s_2}\omega_{\mathbf{q}_3s_3}\right]^{-1/2}\nonumber\\
&\ \ \ \ \times \sum_{NPQ}\sum_{\mu\nu\pi\rho}\sum_{\alpha\beta\gamma\eta}\Phi_{\alpha\beta\gamma\eta}\left(0\mu, N\nu, P\pi, Q\rho\right)\left(M_\mu M_\nu M_\pi M_\rho\right)^{-1/2}\nonumber\\
&\ \ \ \ \times e^{i\mathbf{q}_1\cdot\mathbf{R}\left(N\right)}e^{i\mathbf{q}_2\cdot\mathbf{R}\left(P\right)}e^{i\mathbf{q}_3\cdot\mathbf{R}\left(Q\right)}\nonumber\\
&\ \ \ \ \times w_\alpha\left(\mathbf{q}s,\mu\right)w_\beta\left(\mathbf{q}_1s_1,\nu\right)w_\gamma\left(\mathbf{q}_2s_2,\pi\right)w_\eta\left(\mathbf{q}_3s_3,\rho\right)	\label{4_ph_mat}
\end{align}
\twocolumngrid
respectively, where $w_\alpha\left(\mathbf{q}s,\mu\right)$ is the $\alpha^{\mathrm{th}}$ component of the eigenvector $\mathbf{w}\left(\lambda,\mu\right) = \mathbf{w}\left(\mathbf{q}s,\mu\right)$ for a phonon with wavevector $\mathbf{q}$ and polarization $s$ and for the basis atom $\mu$, and $N_0$ is the number of atoms in the supercell (or equivalently, the number of $\mathbf{q}$-points in the commensurate first BZ). The phonon-isotope scattering probabilities are:
\begin{align}
W^{\mathrm{iso}}_{\lambda\lambda_1} = \frac{\pi\omega_{\lambda}^2}{2N_0}\sum_bg_2\left(b\right)\left|\mathbf{w}\left(\lambda , b\right)\cdot\mathbf{w}^*\left(\lambda_1, b\right)\right|^2\delta\left(\omega_\lambda - \omega_{\lambda_1}\right)
\end{align}
where $g_2\left(b\right) = \left(1/\bar{M}_b^2\right)\sum_af_{ab}\left(M_{ab} - \bar{M}_b\right)^2$  is the mass variance parameter with $f_{ab}$  and $M_{ab}$ being the concentration and mass of the $a^{th}$ isotope of the $b^{th}$ atom, respectively, and $\bar{M}_b$  is the average mass of the $b^{th}$ atom. We find that the momentum-relaxing Umklapp processes are comparable to or stronger than the momentum-conserving normal processes for the four-phonon interactions for all seventeen materials and at all temperatures considered in this work (see SM section S11). As a result, treating the four-phonon scattering within a relaxation time approximation (RTA), i.e., removing the last bracketed term in Eq.~\ref{PBE_4ph}, gives negligible difference compared to the full solution of Eq.~\ref{PBE_4ph}. We have confirmed that (a) solving the full phonon Boltzmann equation, which explicitly differentiates normal and Umklapp three-phonon and four-phonon interactions and (b) solving the partial phonon Boltzmann equation with four-phonon scattering under the RTA by removing the last bracketed term in Eq.~\ref{PBE_4ph} results in negligible difference in $\kappa^{(3+4)}$ at 300 K for all seventeen materials in this study. Therefore, the calculations at all other temperatures presented in this work are obtained by solving the phonon Boltzmann equation with the four-phonon scattering terms treated under the RTA, to reduce the computational cost of these temperature-dependent calculations for the seventeen materials.
 
After solving the phonon Boltzmann equation for $\mathbf{F}_\lambda$, the phonon thermal conductivity is calculated as:
\begin{align}
\kappa_{\alpha\beta} = \sum_{\lambda}C_\lambda v_{\lambda\alpha}F_{\lambda\beta}
\end{align}
where $C_\lambda = \left(1/V\right)k_B\left(\partial n^0_\lambda/\partial T\right)$ is the mode specific heat.
\bibliography{MainManuscript}{}
\bibliographystyle{unsrt}
\end{document}